\documentclass[twocolumn,           
               showpacs,            
               preprintnumbers,     
               aps,                 
               prd,                 
               a4paper,             
               groupedaddress,  
               tightenlines,        
               floats,floatfix      
               ]{revtex4}

\usepackage{graphicx}
\usepackage{latexsym}
\usepackage{amsmath,amssymb}        
\usepackage[draft=false]{hyperref}

\begin{document}



\title{A Bayesian model selection analysis of WMAP3} 
\author{David Parkinson} 
\affiliation{Astronomy Centre, University of Sussex, Brighton BN1 9QH, 
United Kingdom}  
\author{Pia Mukherjee} 
\affiliation{Astronomy Centre, University of Sussex, Brighton BN1 9QH, 
United Kingdom} 
\author{Andrew R.~Liddle} 
\affiliation{Astronomy Centre, University of Sussex, Brighton BN1 9QH, 
United Kingdom} 
\date{\today} 
\pacs{98.80.-k \hfill astro-ph/0605003} 
\preprint{astro-ph/0605003} 
 
 
\begin{abstract} 
We present a Bayesian model selection analysis of WMAP3 data using our
code CosmoNest. We focus on the density perturbation spectral index
$n_{{\rm S}}$ and the tensor-to-scalar ratio $r$, which define the
plane of slow-roll inflationary models. We find that while the
Bayesian evidence supports the conclusion that $n_{{\rm S}} \neq 1$,
the data are not yet powerful enough to do so at a strong or decisive
level. If tensors are assumed absent, the current odds are
approximately 8 to 1 in favour of $n_{{\rm S}} \neq 1$ under our
assumptions, when WMAP3 data is used together with external data
sets. WMAP3 data on its own is unable to distinguish between the two
models. Further, inclusion of $r$ as a parameter weakens the
conclusion against the Harrison--Zel'dovich case ($n_{{\rm S}} = 1$,
$r=0$), albeit in a prior-dependent way.  In appendices we describe
the CosmoNest code in detail, noting its ability to supply posterior
samples as well as to accurately compute the Bayesian evidence.  We
make a first public release of CosmoNest, now available at {\tt
www.cosmonest.org}.
\end{abstract} 
 
\maketitle 
 
 
\section{Introduction} 
 
The recent three-year results from WMAP \cite{wmap3} have provided the 
first firm indications that the spectral index of primordial density 
perturbations, $n_{{\rm S}}$, differs from the Harrison--Zel'dovich 
case $n_{{\rm S}}=1$. The likelihood function in models with varying 
$n_{{\rm S}}$ suggests that $n_{{\rm S}}=1$ is excluded at around 
three to four sigma, in cosmologies with no significant tensor 
contribution to the microwave anisotropies. However, the WMAP team 
stress that their result, based on a chi-squared per degrees of 
freedom argument, needs to be checked using the more sophisticated 
technique of Bayesian model selection \cite{Jeff,MacKay,Gregory}. That 
is the aim of the present paper, building on our previous analysis of 
the WMAP first-year data using our code CosmoNest~\cite{MPL}. 
 
We will consider two different scenarios. The first concerns the 
spectral index alone, under the assumption that there are no 
primordial gravitational waves (parametrized by the tensor-to-scalar 
ratio $r$). As there is presently no indication for gravitational 
waves, this analysis addresses the question of whether $n_{{\rm S}} 
\neq 1$ should be considered part of the standard cosmological 
model. Secondly, we consider the plane of slow-roll inflation models 
parametrized by $n_{{\rm S}}$ and $r$, representing the simplest class 
of inflation models (for an extensive review, see 
Ref.~\cite{LL}). This latter analysis determines the extent to which 
slow-roll inflation models have benefitted from the new data. 
 
\section{Bayesian model selection} 
 
One of the most important classes of statistical problems in science,
and particularly in cosmology, is determining the best fit to data in
the case where the underlying model (i.e.~the set of parameters to be
varied) is unknown. Typically, each parameter represents some physical
effect that may influence the data, but one cannot simply include all
possible physical effects simultaneously, as the data may be
insufficiently constraining and all parameters become undetermined and
biases get introduced \cite{L04}. In the Bayesian framework, the
solution is model selection statistics, which set up a tension
between goodness of fit to the data and model complexity. A model
selection statistic does not care about the preferred values of the
parameters defining the model, but is a property of the model itself,
where here `model' means both a choice of the set of parameters to be
varied and the prior ranges for those parameters.
 
A key application of model selection is to provide a robust criterion
for judging when data requires the addition of new parameters. Many of
the most pressing questions in contemporary cosmology are of this
type, such as whether the dark energy density evolves with redshift,
or whether primordial gravitational waves exist. For the present data
compilation following the WMAP3 announcement, it is the spectral index
which is placed in the most interesting position --- does WMAP3
convincingly exclude the possibility that $n_{{\rm S}}$ is precisely
unity, as conjectured by Harrison and by Zel'dovich \cite{HZ} long
before the inflationary mechanism was discovered? In model selection
terms, does the improved fit that a varying $n_{{\rm S}}$ allows
justify its inclusion as an extra variable fit parameter?
 
One might wonder why we should bother with a model selection analysis 
of a result which a parameter estimation analysis says is already at 
three to four sigma level. The answer is that this significance level 
is exactly where model selection techniques are at their most 
crucial. It has long been recognized in the statistics community that 
Bayesian model selection analyses can give results in contradiction 
with inferences based on `number of sigmas'; this is known as 
Lindley's `paradox' \cite{L} and is nicely summarized by Trotta 
\cite{T}. Basically, Bayesian inference is inconsistent with the idea 
that there is a universal threshold, such as 95\%, beyond which 
results should be seen as definitive; instead such a threshold should 
depend both on the data properties and the prior parameter ranges of 
the models being compared. The Lindley paradox usually manifests 
itself for results with significance in the range two to four sigma 
\cite{T}, which as it happens is exactly where WMAP3 has placed 
$n_{{\rm S}}$. 
 
In a full implementation of Bayesian inference, the key statistic is 
the Bayesian evidence $E$ (also known as the marginalized likelihood), 
which has the literal interpretation of the probability of the data 
given the model \cite{Jeff,MacKay,Gregory}. According to Bayes 
theorem, it therefore updates the prior model probability to the 
posterior model probability. It is simply the average of the 
likelihood over the prior parameter space. Often, the quantity of 
interest is the ratio of evidences of two models $M_1$ and $M_0$, 
called the Bayes factor and denoted $B_{10}$, which indicates how the 
relative model probabilities have been updated by the data. The 
evidence has been exploited in a range of cosmological studies 
\cite{ev,T,MPL}. 
 
Computing the evidence is more challenging than calculating parameter
uncertainties, as it requires knowledge of the likelihood throughout
the prior parameter volume rather than only in the vicinity of its
peak. So far brute force methods such as thermodynamic integration,
though accurate, have proved to be computationally very intensive
\cite{B05}, while approximate information criterion based methods
often lead to results which do not agree and hence can be ambiguous
\cite{L04,M06}. We have recently developed an implementation of an
algorithm due to Skilling known as {\em Nested Sampling}
\cite{Skilling}, which we call CosmoNest \cite{MPL}, which is able to
carry out such calculations efficiently. It is a Monte Carlo method,
but not a Markov chain one. We describe the code extensively in
Ref.~\cite{MPL} and in the appendices of this article.
 
In assessing the significance of a model comparison, a useful guide is
given by the Jeffreys' scale \cite{Jeff}. Labelling as $M_1$ the model
with the higher evidence, it rates \mbox{$\ln B_{10} < 1$} as `not
worth more than a bare mention', $1<\ln B_{10} < 2.5$ as
`substantial', $2.5< \ln B_{10} < 5$ `strong' to `very strong' and
$5<\ln B_{10}$ as `decisive'. Note that $\ln B_{10}=5$ corresponds to
odds of 1 in about 150, and $\ln B_{10}=2.5$ to odds of 1 in 13.

\section{Application to WMAP3} 
 
Throughout we use a data compilation of the WMAP3 TT, TE and EE
anisotropy power spectrum data \cite{wmap3}, together with higher
$\ell$ CMB temperature power spectrum data from ACBAR \cite{ACBAR},
CBI \cite{CBI}, VSA \cite{VSA}, and Boomerang 2003 \cite{boom}, and
also matter power spectrum data from SDSS \cite{SDSS} and 2dFGRS
\cite{2df}. Following the approach of Ref.~\cite{Peiris}, we use the
updated beam error module, and do not marginalize over the amplitude
of SZ fluctuations. For the higher $\ell$ CMB data, we neglect those
bands that overlap in $\ell$ range with WMAP (as in
Ref.~\cite{wmap3}), so that they can be treated as independent
measurements.
 
The prior ranges for the other parameters were chosen as in 
Ref.~\cite{MPL}: $0.018 \le \Omega_{{\rm b}} h^2 \le 0.032$, $0.04 \le 
\Omega_{{\rm cdm}} h^2 \le 0.16$, $0.98 \le \Theta \le 1.1$, $0 \le 
\tau \le 0.5$, and $2.6 \le \ln(A_{{\rm s}} \times 10^{10}) \le 
4.2$. Here $\Theta$ is a measure of the sound horizon at decoupling, 
and the other symbols have their usual meaning. 
 
When we quote Bayes factors, model $M_0$ is always taken to be the 
Harrison--Zel'dovich case. We normalize to this case, which means
positive numbers indicate models preferred against this case.
 
\subsection{The spectral index} 
 
For the spectral index $n_{{\rm S}}$, we will throughout assume a
prior range $0.8 < n_{{\rm S}} < 1.2$, as in Refs.~\cite{B05,MPL}.
The model selection results presented for $n_{{\rm S}}$ must therefore
be understood in light of this prior. As the allowed regions are well
contained within this prior, it is trivial to recompute the Bayes
factor if this range is extended; e.g.~if it is doubled then $\ln
B_{10}$ is reduced by $\ln 2 \simeq 0.7$.

\begin{figure}[t]
\includegraphics[width=0.9 \linewidth]{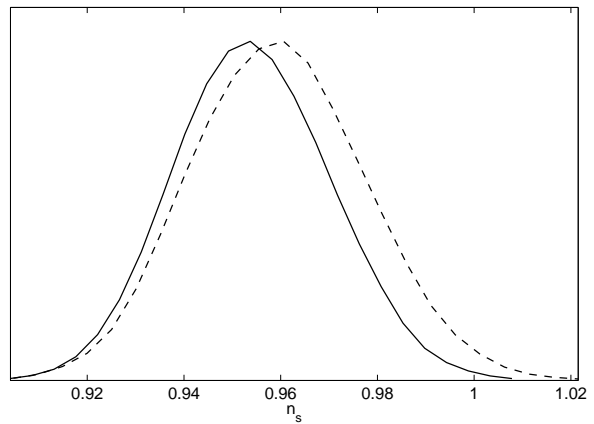}
\caption{Marginalized likelihood of $n_{{\rm S}}$ for WMAP alone 
(dashed) and WMAP+all (solid), obtained using CosmoNest.}
\label{fig:ns}
\end{figure}

Qualitative understanding of our results can be obtained from
studying the marginalized distributions for $n_{{\rm S}}$, shown in
Fig.~\ref{fig:ns}, obtained using CosmoNest. For reasons explained
later, we did not include marginalization over the SZ effect in
obtaining these likelihoods, which shifts them somewhat towards
$n_{{\rm S}} = 1$ as compared to those of the WMAP team \cite{wmap3}.
A rapid guide to the expected result can be obtained by employing a
gaussian approximation to the marginalized posterior distribution for
$n_{{\rm S}}$. As shown by Trotta \cite{T}, the Bayes factor can be
computed in this approximation as a function of $\lambda$, being the
`number of sigmas' of the putative detection, and the information
content $I \equiv \log_{10} (\Delta n_{{\rm S}}/\hat{\sigma})$ which
measures the reduction of the allowed parameter volume between the
prior and posterior (where $\Delta n_{{\rm S}}$ is the prior width and
$\hat{\sigma}$ the standard deviation of the posterior).  For WMAP3
alone and our choice of prior $\lambda \simeq 2.3$ and $I \simeq 1.4$.
Employing Eq.~(18) of Ref.~\cite{T} yields the estimate $\ln B_{10} =
0.4$, i.e.~the varying $n_{{\rm S}}$ model is preferred but only very
mildly. However, this expression assumes that the gaussian form holds
quite far into its tail, which may not be valid, and so we proceed to
results from the full numerical calculation (which we anyway had to do
to obtain these marginalized likelihoods as a by-product, as explained
in Appendix~B).
 
Using CosmoNest, we ran the Harrison--Zel'dovich (HZ) and spectral
index cases ($n_{{\rm S}}$) to find the difference in evidence for
WMAP3, with and without the external CMB and large-scale structure
data sets.  We found that using WMAP alone, $\ln B_{10} = 0.34$, in
agreement with the estimate above.  However, when the extra data sets
are included, $\ln B_{10} = 1.99$, which is substantial but not strong
evidence for the necessity of $n_{{\rm S}}$ as an extra parameter. The
results are given in Table \ref{table1}.

The difference of 1.65 in $\ln E$ between the two datasets can be
understood simply from the marginalized likelihoods shown in
Fig.~\ref{fig:ns}.  Although the curves are similar near the peaks,
and the maximum likelihood value has shifted only by about $0.005$,
the difference in the mean and more importantly the variance have a
large effect in the tails. The probability of the $n_{{\rm S}}=1$
value is about 4 times smaller in the WMAP+all case, which would, all
else being considered equal, translate into a change of $1.4$ in
$\ln E$ explaining most of the difference. Nevertheless, this shift is
not particularly significant on the Jeffreys' scale.

\subsection{The inflationary plane} 
 
A non-zero value of $n_{{\rm S}}-1$ is commonly interpreted as a 
strong indication in favour of inflation. However slow-roll 
inflationary models predict not just a non-zero $n_{{\rm S}}-1$, but 
also a non-zero value for the tensor-to-scalar ratio $r$. A proper 
model comparison motivating inflation should therefore examine not the 
spectral index model but the two-parameter extension of HZ into the 
$n_{{\rm S}}$--$r$ plane. There is no simple way to make an
estimate of the outcome in this case.
 
At this point we run into the issue of the choice of prior for
$r$. While it is uncontroversial to choose a uniform prior for
$n_{{\rm S}}$, whose value is more or less known, $r$ is instead a
parameter whose order-of-magnitude is currently unknown (sometimes
called a `scale' parameter). We will consider two possibilities. The
first, Case 1, is that $r$ has a uniform prior in the range $[0,1]$,
which is the assumption used by Spergel et al.~\cite{wmap3} for
parameter estimation. The second, Case 2, considers the Jeffreys'
prior which states that for scale parameters the prior should be
uniform in $\ln(r)$ rather than $r$. For the problem to be well-defined
this needs to be cut off at both ends. We use the same upper limit,
and as a lower limit take the smallest conceivable inflation scale of
the electroweak scale, which would yield $r \sim \rho^{1/2}/m_{{\rm
Pl}}^2 \sim 10^{-34}$ (where $\rho$ is the energy density). The prior
range is therefore $-80 \leq \ln r \leq 0$. Results are shown in
Table~\ref{table1}.

\begin{table}
\centering
\begin{tabular}{|c|c|c|}
\hline
Datasets & Model & $\ln B_{10}$\\\hline
WMAP only & HZ & $0.0 $ \\
 & $n_{{\rm S}}$ & $0.34  \pm 0.26$ \\   \hline
 WMAP + all & HZ & $0.0 $ \\
  & $n_{{\rm S}}$ & $1.99 \pm 0.26$ \\
  & $n_{{\rm S}}$ + $r$ (uniform prior) & $-1.45 \pm 0.45$ \\
& $n_{{\rm S}}$ + $r$ (log prior) & $ 1.90 \pm 0.24$  \\ \hline
\end{tabular}
\caption{\label{table1} Evidence differences for the different models and 
different data sets, as discussed in the text.}
\end{table}

For Case 1, CosmoNest calculations indicate $\ln B_{10}$ of
$-1.45$. The large amount of unused prior parameter space in the
$n_{{\rm S}}$--$r$ plane means that this model is somewhat disfavoured
as compared to HZ.
 
For Case 2, a calculation is not in fact really necessary, since the
vast majority of the prior space lies in the region where $r$ is
observationally negligible, and hence generates the same likelihood as
a model where the spectral index alone varies. We confirm this
explicitly by computing evidence over a limited range $-5 < \ln r < 0$
and extrapolating the result down to $\ln r =-80$.
 
We conclude therefore that the evidence in the inflationary plane does
carry significant prior dependence, bracketted by the values we have
found under Case 1 and Case 2. Given the present shape of the
likelihood, the evidence for the inflation model will not be as large
as for the spectral index model under any prior choice, and may be
significantly less. For a uniform prior on $r$, the inflation model is
actually rated below Harrison--Zel'dovich.

\subsection{Systematic effects}

The evidence computation we have described takes into account only
statistical uncertainties. However one should also consider the
possible effect of systematic uncertainties, and there are some
indications that these are present at a level which would have some
impact on our conclusions, despite the very careful job that the WMAP
team have done. We highlight some of these issues here.

There is some effect from the precise choice of dataset used.  All the
dataset combinations quoted in Ref.~\cite{wmap3} give very similar
constraints on $n_{{\rm S}}$, though none corresponds precisely to the
dataset compilation we are employing. Curiously though, the dataset
WMAP+all on the LAMBDA archive,\footnote{{\tt
http://lambda.gsfc.nasa.gov}} which adds two supernovae datasets to
our compilation, gives an $n_{{\rm S}}$ value about one-sigma lower
than any other dataset quoted, which would be expected to lead to a
stronger result for the Bayes factor.  However it is puzzling that
this data compilation gives a lower $n_{{\rm S}}$ (and optical depth
$\tau$) than do any of the separate datasets from which it is
compiled.

There is some uncertainty in how to treat the Sunyaev--Zel'dovich
effect and gravitational lensing. The WMAP team allow only for the
former, while Lewis has argued \cite{Lewis} that the two effects are
of the same order, and nearly cancel, at least as regards their effect
on $n_{{\rm S}}$, and that it is better to ignore both than to include
only one. Accordingly we have not included the SZ correction, which
increases $n_{{\rm S}}$ as compared to the WMAP3 analysis.

Another subtlety concerns the modelling of the beam. As discussed in
Ref.~\cite{Peiris} there are different options for doing this, which
appear to have a slight effect on the constraint on $n_{{\rm S}}$. We
have followed the procedure described in that paper, rather than that
of the main WMAP3 papers \cite{wmap3}.

Yet more uncertainty surrounds the modelling of the recombination
process.  According to Ref.~\cite{DG}, inclusion of additional
two-photon decays leads to significant differences as compared to the
standard RECFAST treatment used in the WMAP papers. If confirmed, this
is perhaps not too important for WMAP, but would certainly matter at
Planck sensitivity (Antony Lewis, private communication).

Also, the reionization optical depth $\tau$ and $n_{{\rm S}}$ are
correlated. The constraint on $\tau$ comes mainly from the estimate of
the power in the low $\ell$ multipoles of CMB
polarization. Substantial foregrounds are present in polarization
data, so that their removal using just the frequency information
gathered by WMAP can be tricky. Foreground subtraction uncertainties
could therefore affect $\tau$ and hence $n_{{\rm S}}$.

Finally, we note that the inclusion of Lyman alpha power spectrum data
(not used in the WMAP3 papers) seems to have a marginally significant
effect.  According to the analysis of Ref.~\cite{SSM}, inclusion of
this data shifts $n_{{\rm S}}$ upwards by around one-sigma while
leaving the uncertainty unchanged. Similar results are obtained in
Ref.~\cite{VHL} though the trend is less clear as they round their
quoted results at the second decimal place.

While individually none of the above would have a very major effect on
model selection conclusions, that there are so many clearly urges
caution in interpretting a result whose statistical significance
remains rather marginal.

\section{Conclusions}

We have carried out a Bayesian model selection analysis of WMAP3 data,
as advocated by the WMAP team. We have found that WMAP3 data do indeed
give support for a varying spectral index when combined with other
data, with the Bayes factor compared to the Harrison--Zel'dovich
spectrum being approximately $\ln B_{10} = 2$. According to the
Jeffreys' scale, this should be regarded as significant, but neither
strong nor decisive. It corresponds to probabilistic odds of about 8
to 1 against the Harrison--Zel'dovich model (i.e.~the chance that
$n_{{\rm S}}$ is equal to one is about that of tossing a coin three
times and them all being heads). WMAP3 alone does not provide any
discrimination between the models.

In computing our numbers, we have assumed throughout that the prior
model probabilities are equal, so that models are regarded as equally
likely before the data came along. Anyone who prefers to make an
alternative assumption is welcome to do so, and can readily follow the
consequences using the evidence numbers we have supplied. For
instance, a perfectly plausible standpoint might be that since
inflation is a physical model, its predictions should be taken more
seriously than pure HZ which is motivated only by symmetry
considerations. Hence its prior model probability should be greater,
perhaps tipping the post-data odds decisively against HZ. Readers are
quite welcome to take that viewpoint, but should bear in mind that
their conclusion then derives from a mixture of the data and their
prior prejudice. From the data {\em alone}, the situation remains to
be decisively resolved.

In a companion paper \cite{PLMP}, we forecast the abilities of the
Planck satellite to resolve this situation, in light of the WMAP3
results.


\begin{acknowledgments}

The authors were supported by PPARC. We thank Mike Hobson, Martin
Kunz, Antony Lewis, C\'edric Pahud, Hiranya Peiris, Douglas Scott,
John Skilling, and Roberto Trotta for helpful discussions. We
acknowledge use of the UK National Cosmology Supercomputer (COSMOS)
funded by Silicon Graphics, Intel, HEFCE and PPARC.
\end{acknowledgments}


\appendix

\section{The Nested Sampling Algorithm}

\begin{figure}[t]
\includegraphics[width=0.9 \linewidth]{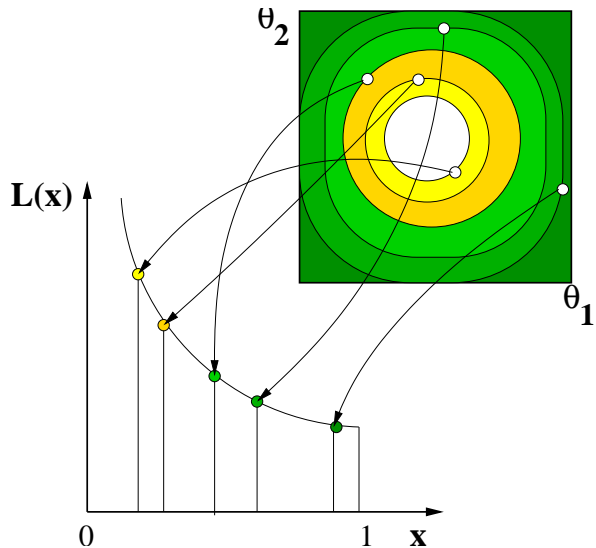}
\caption{Schematic of the Nested Sampling algorithm.}
\label{fig:nested}
\end{figure}

Our implementation of the Nested Sampling algorithm is described in
Ref.~\cite{MPL}.  To summarize, the algorithm (as first developed in
Ref.~\cite{Skilling}) recasts the problem of calculating the evidence
as a one-dimensional integral in terms of the remaining prior mass
$X$, where $dX = P(\theta|M)d\theta$.  So the integral is transformed
\begin{equation}
E = \int L(\bar\theta) p(\bar\theta) d\bar\theta ~~~\rightarrow~~~ E =
\int L(X)dX 
\end{equation}
where $L(X)$ is the likelihood $P(D|\theta,M)$. The algorithm samples
the prior a large number of times, assigning a `prior mass'
probabilistically to each sample.  The samples are ordered by
likelihood, and the integration follows as the sum of the sequence,
\begin{equation}
E = \sum_{j=1}^m E_j\,, \quad E_j=\frac{L_j}{2}(X_{j-1}-X_{j+1}) \,.
\end{equation}
The scheme is illustrated in Figure \ref{fig:nested}.

In order to compute the integral accurately the prior mass is
logarithmically sampled. We start by randomly placing $N$ points
uniformly in the prior space, where in a typical cosmological
application $N \sim 300$. We then iteratively discard the lowest
likelihood point $L_j$, replacing it with a new point uniformly
sampled from the remaining prior mass (i.e.~with likelihood $>$
$L_j$).  Each time a point is discarded the prior mass remaining $X_j$
shrinks by a factor that is known probabilistically, and the evidence
is incremented accordingly. In this way the algorithm works its way
towards the higher likelihood regions.

As the remaining prior mass shrinks by orders of magnitude, the
challenging part is to find an efficient way to draw new points from
the remaining prior volume. We do this by using the $N-1$ remaining
points at each stage to define an ellipsoid that encompasses the
extremes of the points and is aligned with their principal axes. The
ellipsoid is expanded by a constant enlargement factor, in order to
allow for the iso-likelihood contours not being exactly elliptical, as
well as to take in the edges.  New points are then selected uniformly
within the expanded ellipse until one has a likelihood exceeding the
old minimum.

The process is terminated when the integral has been computed to
desired accuracy (see Ref.~\cite{MPL}). In the end the evidence
contributed by the $N-1$ points remaining is added to the accumulated
evidence.

The method is general, and the effects of topology and dimensionality
are implicitly built into it.

\section{CosmoNest}

The Cosmological Monte Carlo code (CosmoMC) developed by Lewis and
Bridle \cite{Lewis:2002} was created to perform an exploration of the
cosmological parameter space, through the Monte Carlo Markov Chain
process (for an overview of MCMC methods see Ref.~\cite{MCMC}). While
it is most commonly used with the Metropolis--Hastings algorithm,
other sampling algorithms (such as Gibbs Sampling and Slice Sampling)
can easily be implemented. The Nested Sampling algorithm can be
considered as just another Monte Carlo sampling algorithm.  The
important difference is that the generation of a chain, which in this
case is not a Markov chain, is ancillary to its primary purpose of
calculating the evidence accurately.

The Cosmological Nested Sampling code (CosmoNest) we have developed is
an additional module that works as part of CosmoMC.

\subsection{Evidence evaluation}

CosmoMC has a `memory' of only one point: the algorithm needs only to
know where it is in order to decide where to go next. CosmoNest needs
to know about the point it is discarding, but must also hold in its
memory all the other $N-1$ live points, as well as knowing how far
through the prior mass $X$ it has progressed and what value of the
evidence ($E$) it has accumulated. The output of a CosmoNest run
consists of the set of discarded minimum likelihood points, along with
their $X$ value, their likelihood, and total accumulated evidence to
that point.

CosmoMC runs multiple chains for two purposes: increasing the speed of
generating samples, and as a way of estimating the extent to which the
chains have explored the parameter space (the Gelman--Rubin
statistic). Here we run multiple iterations of CosmoNest to obtain an
estimate of the uncertainty in the computed evidence.

\subsection{Posterior samples} 
 
The sequence of discarded points from the Nested Sampling process is
similar to the Markov chain produced by an MCMC process with one
important difference: the MCMC points are sampled from the posterior
whereas the Nested Sampling points are sampled from the prior with a
known distribution in $X$. With the appropriate weightings, the
`chain' of discarded points (distributed uniformly in $\ln X$) plus
the remaining live points (distributed uniformly in $X$ within the
remaining volume) can be used to construct the posterior probability
distribution of the parameters, as outlined in Ref.~\cite{Skilling}.
 
\begin{figure}[t] 
\includegraphics[width=0.9 \linewidth]{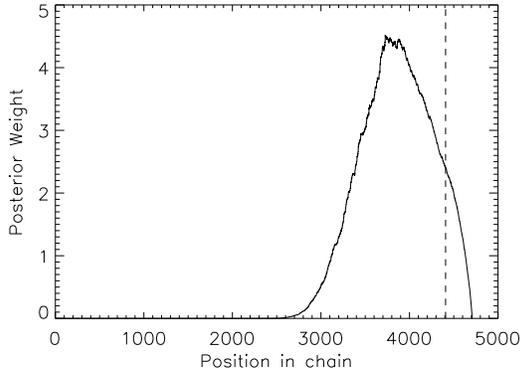} 
\caption{The posterior weights $p_i$ assigned to each point in one of
our HZ runs. The $x$-axis is the element number in the chain, and the
vertical dashed line indicates where the live points start to be
used. This transition could be shifted to the right by running the
code for longer so as to generate a longer chain of discarded points.}
\label{fig:weights} 
\end{figure} 

To summarize, from Bayes' theorem 
\begin{equation} 
p(\theta|D) = \frac{L(D|\theta)\pi(\theta)}{E(D)} \,, 
\end{equation} 
where $p(\theta|D)$ is the posterior probability of a parameter point
$\theta$ given data $D$, $L$ is the likelihood and $\pi$ the prior. So
for an element $i$ in the chain of discarded points, the posterior
weighting is
\begin{equation} 
p_i = \frac{L_i w_i}{E}\,, 
\end{equation} 
where $w_i = \frac12 (X_{i-1} - X_{i+1})$ is the prior mass associated
with that particular point. The $N-1$ points finally remaining also
need to be included to avoid undersampling the centre of the
distribution. They are taken as uniformly sampling the remainder of
the prior space.

\begin{figure}[t] 
\includegraphics[width=0.65 \linewidth]{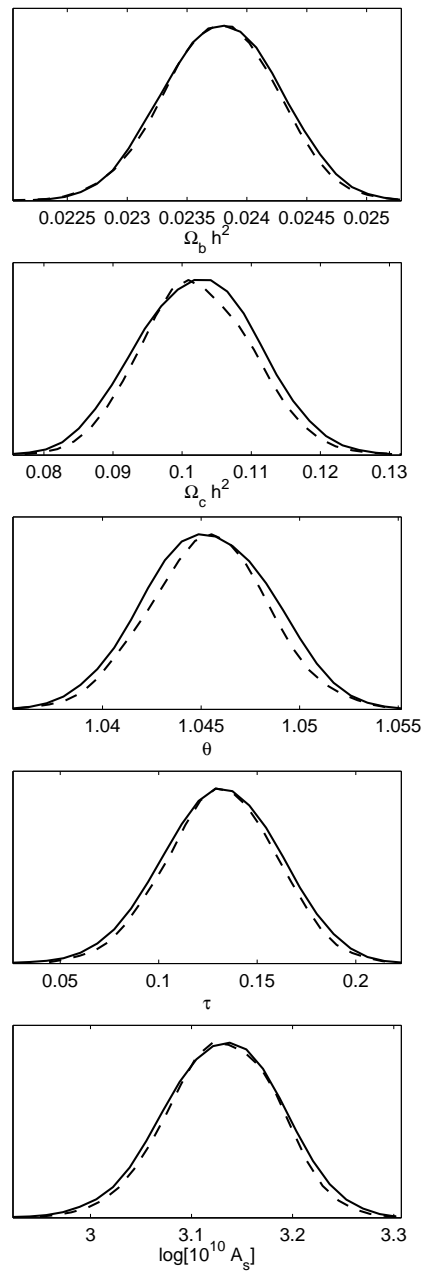} 
\caption{Posterior samples from Nested Sampling (solid) compared to 
MCMC (dashed), for a $\Lambda$CDM HZ model using WMAP3 data only.} 
\label{fig:posterior} 
\end{figure} 

Figure~\ref{fig:weights} shows as an example of the posterior weights
assigned in a particular run. The early points have negligible weight
as their likelihood is low, and the late ones because the prior mass
$w_i$ per point becomes small. We see that in this case the live
points have to be included to properly sample the centre of the
distribution. The fractional contribution from live points can be
reduced by running the code for longer.  The structure of these
weights should be contrasted with Metropolis--Hastings where all
samples have integer weights (values greater than one accruing when
new samples are rejected and instead the original sample duplicated).
 
Using this method we can reconstruct the posterior samples and compare
to similar results from standard Metropolis--Hastings MCMC. We
illustrate this in Fig.~\ref{fig:posterior}. Posteriors obtained from
the two methods are in good agreement.

\subsection{The information}

The information $H$ is defined as (minus) the logarithm of the amount
the posterior is compressed inside the prior \cite{Skilling}. It is
easy to compute from the posterior samples once the evidence has been
calculated:
\begin{equation}
H \equiv \int \ln \left(\frac{dP}{dX}\right) dP = \int \frac{E}{L}
\ln \left(\frac{E}{L}\right)  dX  \,.
\end{equation}
The uncertainty on a single estimate of the evidence is dominated by
the Poisson uncertainty in the number of steps (replacements) to reach
the bulk of the posterior. This is given by $\sigma^2(E) \approx
H\ln[(N+1)/N]$. For the priors we have been considering $H \simeq
10$, and given our choice of $N$, this uncertainty turns out to be
0.15 to 0.2.

\subsection{Public code release} 
 
The CosmoNest code is now freely available for public use, and can be
downloaded from \verb+www.cosmonest.org+. Its use requires a working
installation of the CosmoMC package of Lewis and Bridle
\cite{Lewis:2002}. 
 


\end{document}